\documentclass[conference]{IEEEtran}

\usepackage{cite}
\usepackage{color}
\usepackage{multirow}
\usepackage{booktabs,multirow,array}
\usepackage[table]{xcolor} 
\usepackage{ltxcmds}
\usepackage{footnote}
\usepackage{amsthm}

\usepackage[table]{xcolor}
\usepackage{subfigure}
\usepackage{lipsum}
\usepackage{amssymb} 
\usepackage{pdflscape}
\usepackage{afterpage}
\ifCLASSINFOpdf
\usepackage[pdftex]{graphicx}
\else
  \usepackage[dvips]{graphicx}
\fi

\begin{document}
\title{Design and Development of Automated Threat Hunting in Industrial Control Systems}

\author{\IEEEauthorblockN{Masumi Arafune$^{\ast}$, Sidharth Rajalakshmi$^{\ast}$, Luigi Jaldon$^{\ast}$, Zahra Jadidi$^{\ast\mathsection}$, Shantanu Pal$^{\ast}$, \\ Ernest Foo$^{\mathsection}$, Nagarajan Venkatachalam$^{\ast}$}
\IEEEauthorblockA{$^{\ast}$School of Computer Science, Queensland University of Technology, Brisbane, QLD 4000, Australia \\ 
{$^{\mathsection} $School of Information and Communication Technology, Griffith University, Gold Coast Campus, QLD 4222, Australia}\\
{masumi.arafune@qut.edu.au,
sidharth.rajalakshmi@qut.edu.au,
luigi.jaldon@qut.edu.au,} \\
{zahra.jadidi@qut.edu.au, shantanu.pal@qut.edu.au, e.foo@griffith.edu.au, venkat.venkatachalam@qut.edu.au} {}}}




\maketitle


\begin{abstract}
Traditional industrial systems, e.g., power plants, water treatment plants, etc., were built to operate highly isolated and controlled capacity. Recently, Industrial Control Systems (ICSs) have been exposed to the Internet for ease of access and adaptation to advanced technologies. However, it creates security vulnerabilities. Attackers often exploit these vulnerabilities to launch an attack on ICSs. Towards this, threat hunting is performed to proactively monitor the security of ICS networks and protect them against threats that could make the systems malfunction. A threat hunter manually identifies threats and provides a hypothesis based on the available threat intelligence. In this paper, we motivate the gap in lacking research in the automation of threat hunting in ICS networks. We propose an automated extraction of threat intelligence and the generation and validation of a hypothesis. We present an automated threat hunting framework based on threat intelligence provided by the ICS MITRE ATT\&CK framework to automate the tasks. Unlike the existing hunting solutions which are cloud-based, costly and prone to human errors, our solution is a central and open-source implemented using different open-source technologies, e.g., Elasticsearch, Conpot, Metasploit, Web Single Page Application (SPA), and a machine learning analyser. Our results demonstrate that the proposed threat hunting solution can identify the network's attacks and alert a threat hunter with a hypothesis generated based on the techniques, tactics, and procedures (TTPs) from ICS MITRE ATT\&CK. Then, a machine learning classifier automatically predicts the future actions of the attack.
\end{abstract}

\IEEEpeerreviewmaketitle

\section{Introduction}


Traditionally, industrial systems were isolated from external access, and security was not a primary design criterion. Many of today's Industrial Control Systems (ICSs) are exposed to the Internet, creating security vulnerabilities due to the lack of proper security solutions~\cite{b14}. An ICS adversary often practices different actions to exploit these vulnerabilities, pass the border between Information Technology (IT) and Operational Technology (OT) networks and launch a targeted attack against ICS networks. Many organisations use cyber threat hunting to detect hidden intrusions before they cause a significant breach proactively~\cite{b6}. Hunting aims to detect threat actors early in the cyber kill chain by searching for signs of an intrusion and then providing hunting strategies for future use~\cite{assante2015industrial}. While threat hunting in conventional communications networks is not novel, the idea of threat hunting in ICS networks consisting of a combination of IT and OT networks is a new challenge due to the diverse nature of OT networks \cite{b8} \cite{b9} \cite{b10}. This necessitates an automated threat hunting solution that can provide adequate security to monitor and control the operation of ICS networks. 


MITRE’s Adversarial Tactics, Techniques, and Common Knowledge (ATT\&CK) is the model of cyber adversary behaviour~\cite{b32}. It outlines different phases of attacks’ lifecycle in IT and OT networks and the platforms the attackers are known to target. This model helps to expand the knowledge of threat hunters by outlining the tactics, techniques, and procedures (TTPs) which adversaries use to gain access to a system and execute their targeted attacks. It is difficult for adversaries to change their TTP behaviour when they are operating them. Therefore, MITRE ATT\&CK has focused on TTPs. TTPs provided by MITRE ATT\&CK are sorted based on the attack life cycle. This can help threat hunters to generate a hypothesis based on the initial TTPs of an attack and predict other possible techniques across future phases of the attack life cycle. 
MITRE ATT\&CK consists of two categories: (i) MITRE ATT\&CK for IT \cite{b32}, and (ii) MITRE ATT\&CK for ICS (released in 2020) \cite{b33}.
MITRE ATT\&CK for ICS presents adversarial TTPs against OT systems. However, as it has been released recently, the application of this model in threat hunting in ICS networks has not been sufficiently studied. 

In the MITRE ATT\&CK framework, tactics show what objective the attacker wanted to achieve with the compromise. For each tactic, a wide array of techniques that threat actor groups have used are provided by the framework. Furthermore, the ATT\&CK for ICS matrix can help determine what types of data sources are required to detect threats in ICS environments \cite{b4} \cite{b13}  \cite{haddadpajouh2021survey}. Threat hunting is a human-based approach, and it is prone to human errors~\cite{b34} \cite{khalid2021advanced} \cite{pal-wowmom}. Therefore, automation of the threat hunting process can help reduce human errors and increase detection speed.


Unlike the existing threat hunting solutions, which are cloud-based, costly, and IT-focused, this paper is motivated to design and develop a central and open-source automated threat hunting solution for ICS networks. In this paper, we investigate threat hunting using MITRE ATT\&CK for ICS networks. The input data for our threat hunting solution is received from Ethernet-based devices connected to an ICS network. The major contributions of the paper are as follows:

\begin{itemize}
\item We design a central and open-source framework for automated threat hunting in ICS networks.


\item We provide a detailed proof of concept implementation of the proposed automated threat hunting solution. The implementation consists of the following two phases: 

\begin{itemize}
    \item \textit{Automatic detection of adversarial TTPs:} the first step in automating threat hunting is developed from the fact that attackers mostly use common adversarial TTP’s which are stored in a database or framework specifically developed for more effective threat hunting. To detect TTPs in ICSs, we provide a central threat hunting platform that communicates with MITRE ATT\&CK for ICS framework and automatically detect TTPs.


\item  \textit{Prediction of future TTPs:} a supervised machine learning method is used to provide an automatic analysis of attack TTPs,  generate and validate a hunting hypothesis, and predict the future steps of the detected ICS attack. 
\end{itemize}
\end{itemize}


{\color{black}
To evaluate the accuracy of the automated threat hunting methods, sample training and testing datasets were generated based on MITRE ATT\&CK for ICS to train and test machine learning algorithms for real-world APT attacks. 
}


The rest of this paper is organised as follows. In Section~\ref{sec:RelatedWorks}, we discuss related works. Section~\ref{sec:Architecture} explains the proposed framework of our open-source threat hunting method. ICS dataset generation and the framework components are discussed in Section \ref{ics-data-generation} and Section \ref{central-threat-hunting}. The implemented machine learning-based classification is presented in Section \ref{sec:ML}. In Section \ref{discussion}, we present the achieved results and discuss the major findings. Finally, in Section \ref{sec:Conclusion}, we conclude the paper with future work. 

\section{Related Work}
\label{sec:RelatedWorks}



Several studies examine threat hunting in IT networks \cite{b25} \cite{bhardwaj2019framework} \cite{raju2021survey} \cite{darabian2020multiview}. Yet, there is a need for proactive threat hunting solutions that should be employed in ICS networks due to the growing Advanced Persistent Threats (APTs) against industrial networks. However, this area of research is lacking in recent literature.  Furthermore, the automation of threat hunting is another challenge that has not been adequately addressed even in IT networks. While there are some papers about the automation of hunting in IT networks~\cite{b5, b20, b26, b24, b15}, more research is needed for ICS networks.   

In~\cite{b5}, known APTs are used to generate automatic hypotheses which are then assessed with given probability readings. The automatic hypothesis generation uses an Intrusion Detection System (IDS) formerly known as BRO. The BRO IDS generates alerts with attached hypotheses, the hypotheses are then compared to the detected APT’s to determine whether they are intruding or false positives~\cite{b5}. 


A deep learning stack was developed in \cite{b20} to observe APTs as a multi-vector multi-stage attack that can be captured by using the entire network flow and raw data as input for the detection process. The solution considered outliers, data dimensions, non-linear historical events, and previously unknown attacks. Further, it used different algorithms to perform detection and classification. However, unlike their approach, we focus on automation of threat hunting in ICS networks.

Proposal \cite{b26} developed a threat hunting method for IT networks using a machine learning-based system to detect and predict APT attacks through a holistic approach. The system consists of three main phases: threat detection, alert correlation, and attack predictions. The proposed system is able to capture attacks in a timely fashion. 

Another hunting solution for IT environment is proposed in \cite{b24}. It deployed in an Ubuntu virtual machine for detecting APT tactics through synthesised analysis and data correlation. The framework achieves APT tactic detection through logs, configuration files and previously seen APT tactics, which are taken as inputs, and by using these variables, a ranked list of APT tactics based on completeness is generated. 

DFA-AD is another architecture designed for the detection of APTs by using event correlation techniques \cite{b15}. The system acquires its information through collected and processed network traffic packets. 
It used a three-staged approach.
The first stage consists of four classifiers, each with a different detection scheme to detect all techniques used in various steps of an APT attack. In the second stage, the event correlation modules take all the event outputs from the classifiers and correlate all of them individually to find a caution on APT attack discovery. Finally, in the third stage,  a voting service is used that analyse the correlations and determine a result. Through this voting phase, DFA-AD can lessen the rate of false positives and is able to increase the accuracy of APT detection. 


The above-discussed papers proposed automation of threat hunting in IT networks - with a significant gap in presenting the automated hunting in ICS networks. Nevertheless, threat hunting in the ICS domain lacks sufficient research about feeding the MITRE ATT\&CK framework into threat hunting tasks. Other automated solutions have utilised the MITRE framework to perform threat hunting and detect APTs. For instance, \cite{b6} and \cite{b7} employed MITRE ATT\&CK for Enterprise to detect threats. Proposal~\cite{b7} further used statistical analysis based on MITRE ATT\&CK to learn APT TTPs to predict the future techniques that the adversary may perform. However, unlike our approach, once again, they do not consider the automation of threat hunting in ICS networks.


In \cite{b21}, a framework called ‘Spiking One-Class Anomaly Detection’ based on the evolving ‘Spiking Neural Network Algorithm’ is discussed for APT detections in ICS networks. The algorithm implements the one-class methodology’s innovative application in training a model with exclusive data that characterise the operation of an ICS. It can detect abnormal behaviours and is suitable for applications with massive amounts of data. However, this algorithm did not utilise an application to display the results of the test data.


Another detection method in an ICS, \cite{b22}, used a spatio-temporal association analysis method to detect intrusions in industrial networks. It focused on feature mining and retrieval methods of historical attacks between the features of APT attacks. The proposal used a multi-feature SVM classification detection algorithm to detect abnormal APT attacks. However, unlike our motivation, the solution does not explicitly state which of the many APT groups is attacking or discuss the technique used in the detected attack.

In summary, we note that there are still areas that need further development regarding threat hunting, specifically that it is imperative to incorporate the knowledge of existing APTs. At the same time, there is a need for additional validation of defence mechanisms and to enhance these defences so that they can be easily integrated into ICSs, which is missing in recent literature. Our paper proposes a framework that addresses (i) threat hunting in ICS networks, (ii) applies new MITRE ATT\&CK for ICS framework to the threat hunting process, and (iii) automate the hunting tasks. In addition, our framework can predict possible TTPs that may happen in the future and detect the type of APTs using these TTPs.

\section{Proposed Framework}
\label{sec:Architecture}

In Fig.~\ref{fig-framework}, we illustrate an outline of the proposed framework. 
Recall, while the existing cloud-based platforms may introduce new security issues when sensitive information is sent to the cloud, our threat hunting approach aims to be a centralised and automated solution for ICS networks. 
At first, the data is received from the ICS devices. The red dotted line illustrated in Fig. \ref{fig-framework} shows the core of the centralised automated threat hunting.
Our framework uses MITRE ATT\&CK and automatically generate and validate a hypothesis. Finally, the proposed framework uses a supervised machine learning method to predict the future behaviour of the detected TTPs. 
The functionality of the proposed framework comprises the following steps:

\begin{figure}[t]
\centering
\includegraphics[scale=.6]{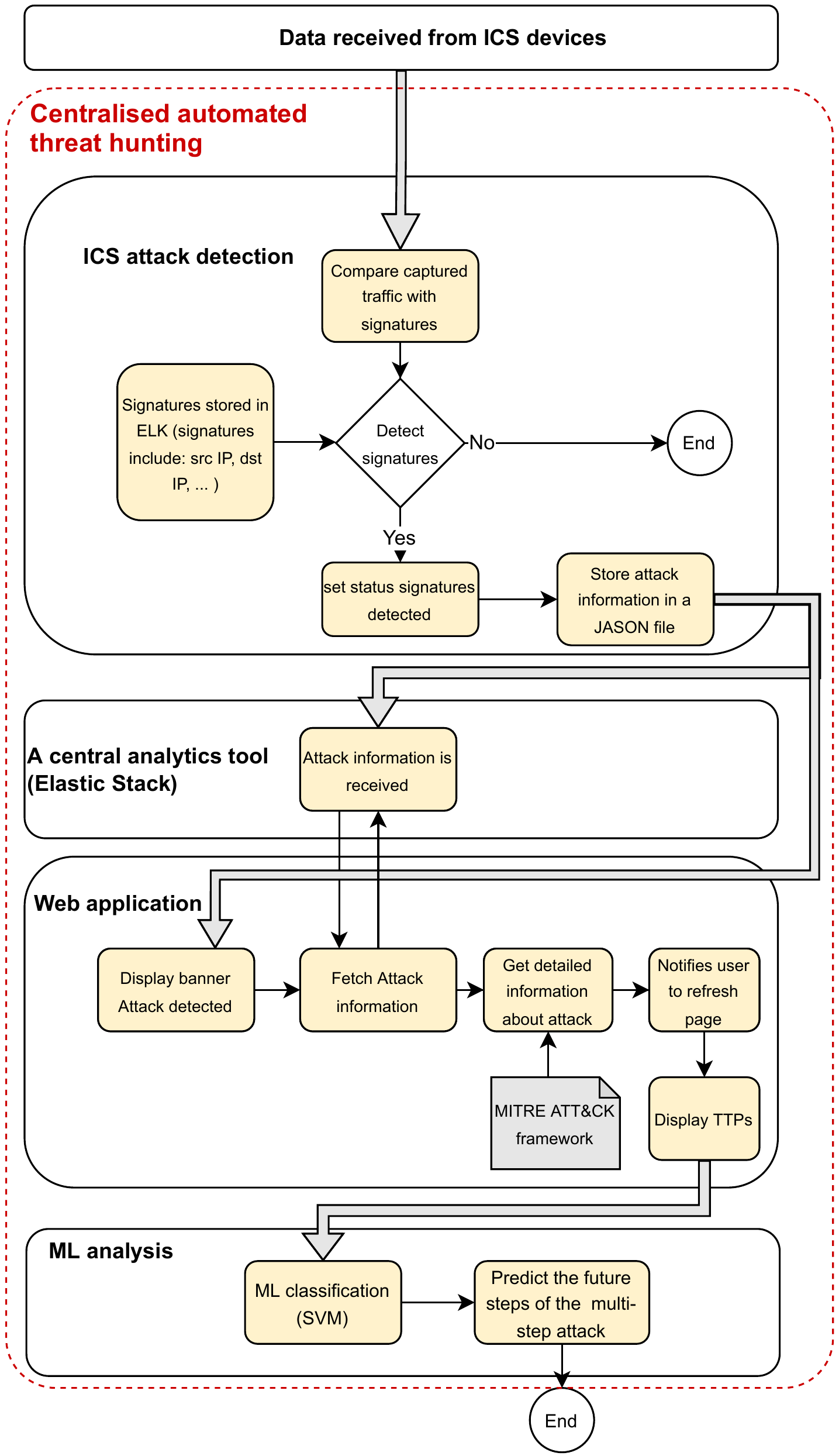}
\caption{The framework of the proposed solution.} 
\label{fig-framework}
\end{figure}

\begin{itemize}
\item ICS data generation: A Conpot honeypot and Metasploit were used to simulate ICS attack techniques. 
\item Central threat hunting platform: The signatures of ICS attacks were generated and stored in a repository to detect attacks. An open-source security analytics tool like Elastic Stack (ELK) was used in this phase to communicate with ICS detection phase and MITRE ATT\&CK framework to detect TTPs. A web Single Page Application (SPA) was used in the web application phase to display the TTPs. 
\item Machine learning analysis: a supervised machine learning approach was used in this phase to detect the possible APTs and predict the future TTPs that can be used by the identified APTs.
\end{itemize}

The proposed framework identifies the attacks in the network and alerts the threat hunter with a hypothesis generated based on the TTPs from MITRE ATT\&CK. 
Open source Elastic-search stores the attack information as well as the TTPs from MITRE ATT\&CK. The web application which threat hunters can utilise for interacting with the platform, get notified of any attacks along with a valid hypothesis.
Note, in this research, an important issue was the generation of dataset, identification and detection of attack based on its signature, generating hypotheses from MITRE ATT\&CK TTPs, and finally, sending an alert to the web application. 
%
The codes of our proposed automated hunting solution can be found in the following repositories  
(Front End\footnote{
https://github.com/massarafune/ICS-threat-hunting-reactapp} and Signature Detection\footnote{https://github.com/mistersiddd/signature-detection-pcap}).
{\color{black} Next, we discuss different phases of the proposed solution.}

\section{ICS Data Generation}
\label{ics-data-generation}
In this phase, we filtered incoming network traffic to look for unique packets which contained attack signatures. These signatures can be used to detect incoming attacks as well as identify what attacks they are and create a hypothesis of potential attack objectives and procedures. In other words, once we identify the type of attack, it is possible to verify the hypothesis based on the MITRE ATT\&CK framework. Therefore, we could identify possible solutions for the given problem if we could identify attack signatures seen in network traffic. Thus, it is possible to create software that automates the detection, identification, and verification of the threat hunting process. We have three automated tasks in our solution: (i) \textit{Detection} (when the attack happens), (ii) \textit{Identification} (what kind of attacks), and (iii) \textit{Verification} (providing possible attack details).

\begin{figure}[t]
\centering
\includegraphics[scale=0.7]{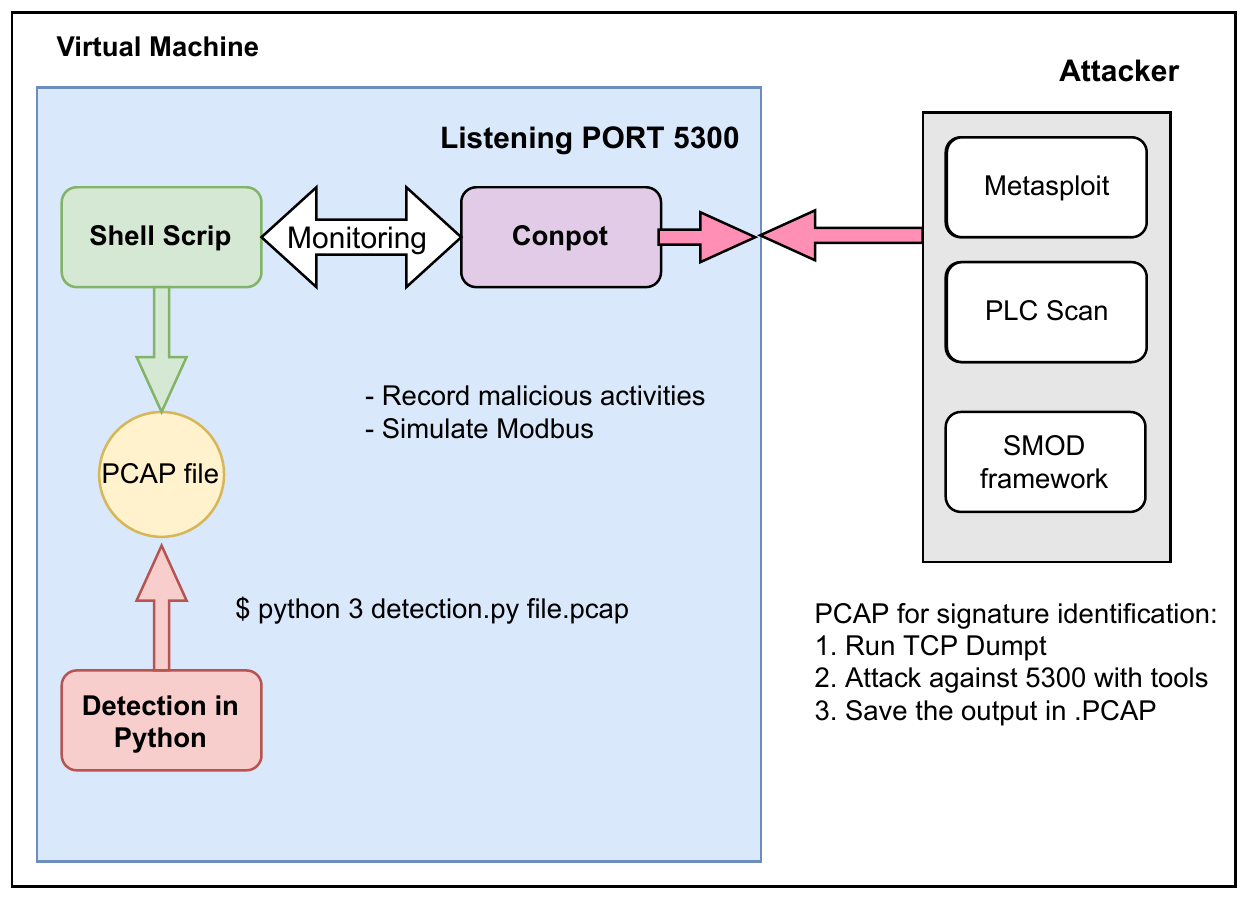}
\caption{The TTP data generation process.} 
\label{ttp-generation}
\end{figure}

The results which will be presented in the paper are provided as a proof of concept for our automated threat hunting solution. 
A few sample sequences of TTPs were generated in this paper to simulate APT behaviour. Modbus-based TTPs are used to explain our solution. The Procedure of creating data is shown in Fig.~\ref{ttp-generation}. It is possible to simulate some of the potential attack scenarios by utilising ICS honeypot (Conpot) and open-source attacking tools e.g., (i) Metsploit framework, (ii) PLC Scan (a lightweight PLC scanner used for ModBus and S7comm protocols), and (iii) Smod framework (a Modbus penetration testing framework).

In our framework, Conpot was installed in a virtual machine, and it is listening to Port 5300, where the honeypot is listening to as a Modbus, serving the Modbus facility as a honeypot. An attacker can send a request to conpot, conpot is listening to so many ports, including FTP, SSH, HTTP. However, we used Port 5300. We listen and monitor port 5300, where conpot and Modbus were running, and if something happened against 5300 on this interface, the shell script will be monitoring all the traffic in enps08 and save it in a PCAP file. Then, we can directly pass the PCAP file to python, python will be run by a shell script. Conpot is continuously listening here, a shell script is not running all the time. It is done manually if we have a PCAP file to dump it so that python can run it as an argument. We have those three to generate the PCAP file.
We created a hypothetical attacking scenario against the honeypot to find and extract attack signatures. 
We made the following assumptions:
\begin{itemize}
  \item Attackers try to scan the target environment to find possible attack vectors. 
  \item Attackers try to identify what type of equipment is installed in the environment. 
  \item Attackers try to identify the UID of installed equipment for further attacks. 
  \item Attackers try to modify the state of the equipment. 
\end{itemize}

Based on the assumptions, we conducted a series of attacks and generated network traffic data that was used to identify attack signatures. 

\section{Central Threat Hunting} 
\label{central-threat-hunting}
A centralised threat hunting platform is used in this phase to receive detected attacks and detect TTPs employed by these attacks. This phase: 

\begin{itemize}
  \item Provides a user interface for users to review attacks. 
  \item Stores attacks history and MITRE ATT\&CK framework. 
  \item Detects and identify incoming attack TTPs instantly. 
\end{itemize}

Therefore, our solution, Threat Hunting Platform, consists of three components, (i) a signature-based detection system to detect and identify incoming attack TTPs, (ii) Elastic-Search to store and provide a feature-rich database function, and (iii) a web application to provide a user interface. 

\subsubsection{ICS TTP Detection}
The signature-based attack detection used in this phase is developed using python because Elastic-Search provides a python module to allow developers to interact with stored data effectively. This system would go through the network traffic and identify malicious payloads based on known attack signatures. Upon identification, the attackers IP, victim’s IP and timestamp (the time when the malicious packet arrives) would be sent to Elastic-Search to be stored. 
The signature detection system detects malicious activities found in the data section of the TCP datagram and identifies attack types by searching for known signatures. Using Scapy, this python script can scan known attack signatures from raw network traffic. Also, detected packets are labelled as malicious. The associated source IP address and destination address, as well as a timestamp of the packet, would be sent to Elastic-Search via Elastic-Search API. 

Since we mapped those signatures and types of TTPs, we could identify employed tactics and techniques for the attack, the python script sends identified TTPs to Elastic-Search along with packet information that is mentioned above. 
Upon detecting malicious activities, the python script sends notifications to the web application so that users can be aware of the TTPs instantly. 

\subsubsection{Central Analytic Tool} 

Elastic-Search provides a wide range of features, including Logstash to import log files from any nodes instantly and Filebeats to sync the database with specific files in a system automatically. These features are helpful to import the MITRE ATT\&CK framework dataset to our solution. Also, considering that we deliver our solutions in a virtual machine, the MITRE ATT\&CK dataset should be available offline. Therefore, we decided to clone the MITRE ATT\&CK framework GitHub repository to our virtual machine so that a user can update its local repository anytime to access up-to-date attack information locally.

\subsubsection{Web Application}
\begin{figure}[t]
\centering
\includegraphics[scale=0.71]{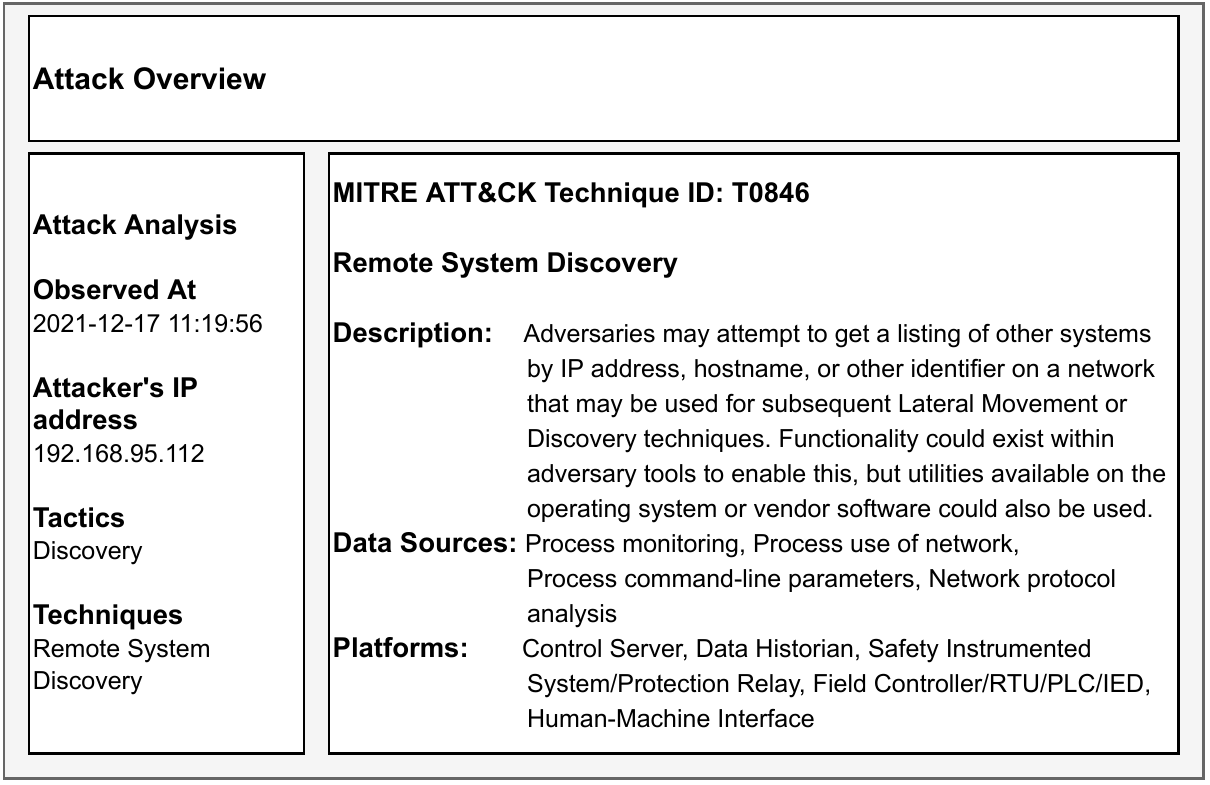}
\caption{Attack information.} 
\label{fig-4}
\end{figure}

A SPA written in React JS (JavaScript) provides the dashboard showing some data in Elastic-Search. 
\begin{figure}[t]
\centering
\includegraphics[scale=0.65]{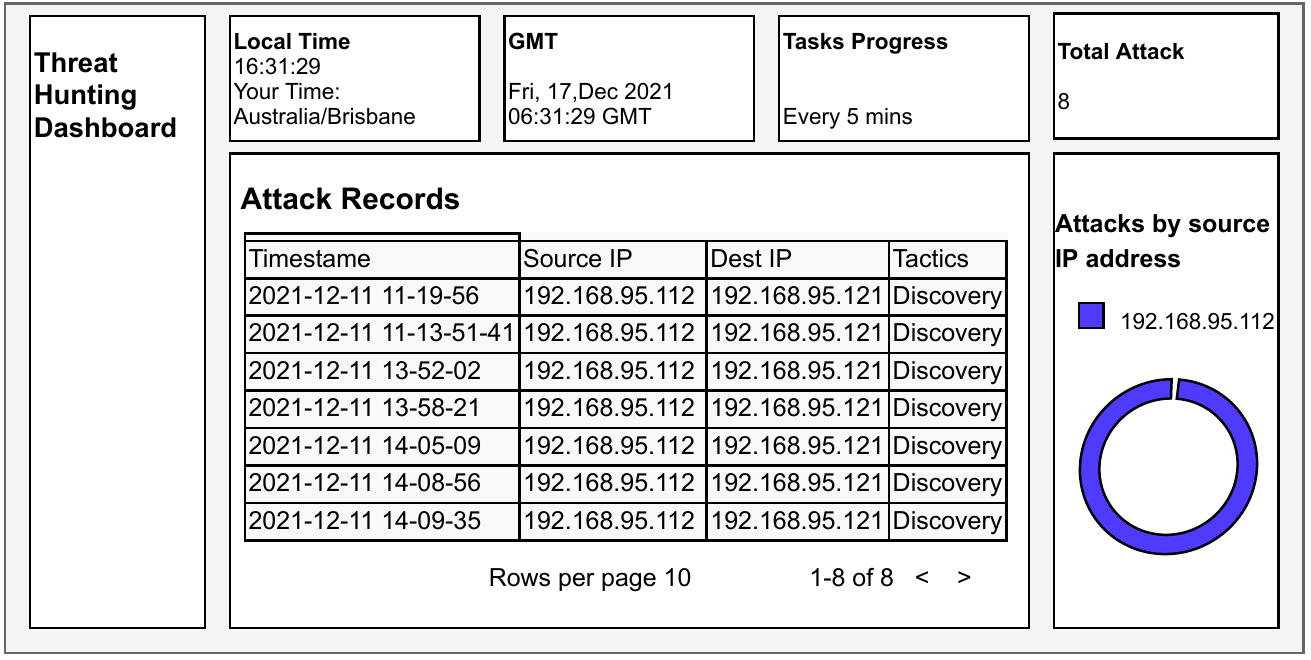}
\caption{Front-end homepage.} 
\label{fig-3}
\end{figure}
As shown in Fig.~\ref{fig-3}, the application provides a list of attack history where users can review attack TTPs with basic information e.g., attackers IP address, victim's IP address, time stamp and type of attacks. 
Users can find detailed information about attacks by looking at a row in the table (Fig.~\ref{fig-3}). As shown in Fig.~\ref{fig-4}, users can investigate the attack TTPs and read the attack description from the MITRE ATT\&CK framework.
This application also provides alert notifications when an attack happens. The web application is designed to closely communicate with other components and display alerts as shown in Fig.~\ref{fig-4}. 


\section{Machine Learning Analysis}
\label{sec:ML}
In this section, we discuss the achieved results and provide a dissuasion on the major findings. The output of our selected TTPs feeds the machine learning analysis phase. In particular, we used a supervised machine learning approach in this phase to detect the APT associated with the TTPs. In this phase: (i) a training dataset was generated using MITRE ATT\&CK Matrix including IT and OT TTPs, and (ii) a supervised support vector machine (SVM) classifier was trained by the MITRE ATT\&CK data to predict possible future APTs.
Next, we discuss them in detail. 

\subsubsection{Dataset Generation}

We generated training and testing datasets based on the MITRE ATT\&CK framework to automate the detection of TTPS through the integration of machine learning, and semantic analysis \cite{b31}. Some samples generated using MITRE ATT\&CK for ICS are displayed in Table~\ref{tab-1}. TTPs used by different APTs are indicated by 1s in the dataset (as shown in Table~\ref{tab-1}). 
 


\begin{table}[t]
\caption{Sample Dataset for ICS matrix.}
\label{tab-1}
\resizebox{\columnwidth}{!}{\begin{tabular}{c c c c c c c c c c c c c}
\hline
\rotatebox{90}{APT Label}&\rotatebox{90}{Initial Access} & \rotatebox{90}{Execution} & \rotatebox{90}{Persistence} & \rotatebox{90}{Privilege Escalation} & \rotatebox{90}{Evasion} & \rotatebox{90}{Discovery} &  \rotatebox{90}{Lateral Movement} & \rotatebox{90}{Collection} & \rotatebox{90}{Command and Control} & \rotatebox{90}{Inhibit Response Function} & \rotatebox{90}{Impair Process Control}& \rotatebox{90}{Impact} \\
\hline
\hline
1 & 1 & 0 & 1 & 0 & 0 & 0 & 0 & 0 & 1 & 0 & 0 & 0 \\
2 & 0 & 1 & 1 & 0 & 1 & 1 & 1  & 0 & 0 & 1 & 0 & 0 \\
3 & 1 & 0 & 0 & 0 & 0 & 0 & 1  & 0 & 0 & 0 & 0 & 0 \\
4 & 0 & 0 & 0 & 0 & 0 & 0 & 0  & 1 & 0 & 0 & 0 & 1 \\
\hline
\end{tabular}
}
\end{table}

\subsubsection{SVM Classifier}
An SVM classifier was implemented and evaluated by the generated datasets. Both MITRE ATT\&CK for Enterprise and MITRE ATT\&CK for ICS were used to train the classifier to be able to detect attacks in both IT and OT sections of an ICS network. 
The SVM classifier could accurately predict the attacking groups of ICS TTPs with an accuracy of 100\%, and it showed 98\% accuracy in detecting IT TTPs. The advantage of this machine learning phase is that when the central hunting phase detects a few TTPs, this SVM classifier predicts all APTs which have these TTPs in common, and the hunter will receive the details of possible future TTPs. 
\section{Discussion}
\label{discussion}
\begin{figure}[t]
\centering
\includegraphics[scale=0.33]{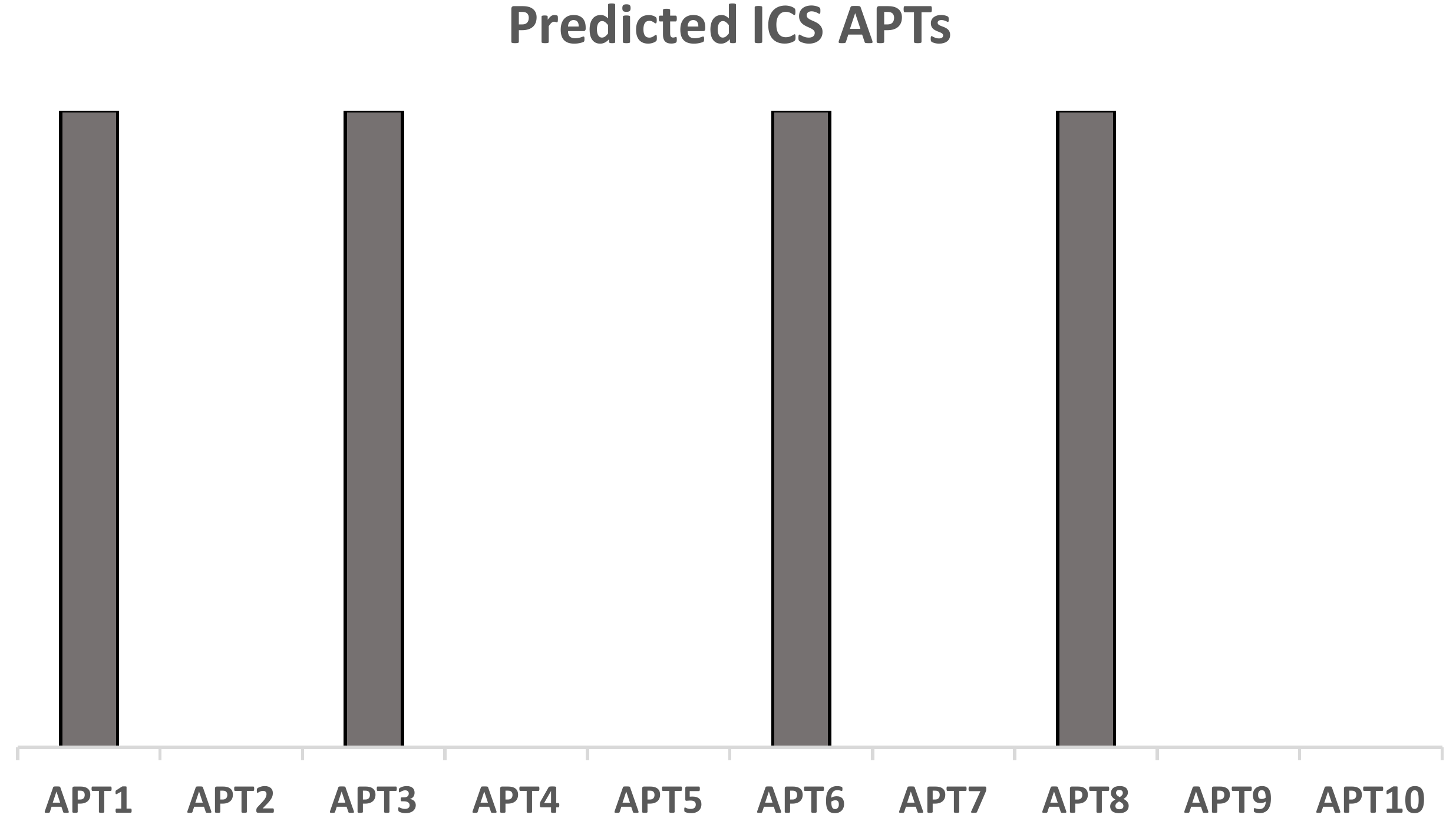}
\caption{Predicting APTs based on TTPs detected in our network.} 
\label{fig-6}
\end{figure}

The proposed framework is of significant use for the ICS due to its novel approach to threat hunting. The solution could be automated in ICS organisations which in turn minimises human errors, time and overall costs in the threat hunting process. Further, the framework can provide detailed information regarding the attacks based on the MITRE ATT\&CK framework, thereby making it useful for ICS industries to interpret the attacks in an extensive manner. This type of application is also rarely found in the industry of ICS. 
Using the dataset in Table~\ref{tab-1}, an SVM classifier was used to analyse the TTPs detected in the network and identify the APT groups that used the same TTPs. The SVM classifier can accurately predict the attacking groups in both IT and OT Matrix. Since the high accuracy was confirmed, it was implemented using Python and it posts the predicted groups and detected TTPs to Elastic-Search. The index is requested by a server-side application developed using the Express JS framework. An additional single page application was developed using React JS, which retrieved this data from the server to display the detected TTPs and predicted group. 
As shown in Fig.~\ref{fig-6}, a bar chart is provided for the users to see the adversaries detected in their systems. The chart can display multiple groups detected when specific TTPs are given as input. It also allows users to find out more about the groups detected as it will take them to the group's details on the MITRE framework website. 
In Fig.~\ref{fig-6}, react application predicts multiple APT attacks in our networks using the two TTPs detected by the central threat hunting phase.

\section{Conclusion and Future Work}
\label{sec:Conclusion}
Threat hunting aims to detect threat actors early in the cyber kill chain by searching for signs of intrusions and then providing hunting strategies for future use. In response to concerns about hidden intrusions, several cloud-based hunting platforms have been proposed for IT networks that facilitate monitoring of distributed devices and help detect ATT\&CK TTPs associated with detected anomalies. In this paper, we have presented a comprehensive approach that can be used by industrial companies to implement central and automated threat hunting in ICS networks. 
Our proposed open-source solution uses MITRE ATT\&CK for ICS to detect TTPs and then, automatically find the APTs associated with the input TTPs. Our method can potentially detect malicious activities more quickly and accurately. The machine learning-based analyser in our solution can predict the future steps of the identified TTPs. It is also used to automate decision making tasks e.g., the verification of the hypothesis. Note, this paper studied an automated threat hunting process in ICS networks as a proof of concept.
 In the future, we intend to explore how graph analysis can help our framework to improve threat hunting. 
\section*{Acknowledgment}
The authors acknowledge the support of the Commonwealth of Australia and Cyber Security Research Centre Limited.

\ifCLASSOPTIONcaptionsoff
  \newpage
\fi

\bibliographystyle{IEEEtran}
\bibliography{bare_jrnl}

\end{document}